\documentclass[fleqn,usenatbib]{mnras}

\usepackage{newtxtext,newtxmath}

\usepackage[T1]{fontenc}
\usepackage{ae,aecompl}
\usepackage{float}
\usepackage{changes}

\usepackage{graphicx}	
\usepackage{amsmath}	
\usepackage{ulem}
\usepackage{amssymb}	

\title[Gravitational waves from SGRs/AXPs as fast-spinning WDs]
{Gravitational waves from SGRs and AXPs as fast-spinning white dwarfs}
\author[M. F. Sousa et al.]{
Manoel F. Sousa,$^{1}$\thanks{E-mail: manoel.sousa@inpe.br}
Jaziel G. Coelho,$^{1,2}$\thanks{E-mail: jazielcoelho@utfpr.edu.br} and
Jos\'e C. N. de Araujo$^{1}$\thanks{E-mail: jcarlos.dearaujo@inpe.br}
\\
$^{1}$Divis\~{a}o de Astrof\'{i}sica, Instituto Nacional de Pesquisas Espaciais, Avenida dos Astronautas 1758, S\~{a}o Jos\'{e} dos Campos, SP 12227-010, Brazil \\
$^{2}$Departamento de F\'isica, Universidade Tecnol\'ogica Federal do Paran\'a, 85884-000 Medianeira, PR, Brazil\\
}

\date{Accepted XXX. Received YYY; in original form ZZZ}

\pubyear{2020}

\begin{document}
\label{firstpage}
\pagerange{\pageref{firstpage}--\pageref{lastpage}}
\maketitle

\begin{abstract}
In our previous article we have explored the continuous gravitational waves (GWs) emitted from rotating magnetized white dwarfs (WDs) and their detectability by the planned GW detectors such as LISA, DECIGO and BBO. Here, GWs emission due to magnetic deformation mechanism is applied for Soft Gamma Repeaters (SGRs) and Anomalous X-Ray Pulsars(AXPs), described as fast-spinning and magnetized WDs. Such emission is caused by the asymmetry around the rotation axis of the star generated by its own intense magnetic field. Thus, for the first time in the literature, it is estimated the GWs counterpart for SGRs/AXPs described as WD pulsars. We find that some SGRs/AXPs can be observed by the space detectors BBO and DECIGO. In particular, 1E 1547.0-5408 and SGR 1806-20 could be detected in 1 year of observation, whereas SGR 1900+14, CXOU J171405.7-381031, Swift J1834.9-0846, SGR 1627-41, PSR J1622-4950, SGR J1745-2900 and SGR 1935+2154 could be observed with a 5-year observation time. The sources XTE J1810-197, SGR 0501+4516 and 1E 1048.1-5937 could also be seen by BBO and DECIGO if these objects have $M_{WD} \lesssim 1.3 M_{\odot }$ and $M_{WD} \lesssim 1.2 M_{\odot }$, respectively.
We also found that SGRs/AXPs as highly magnetized neutron stars are far below the sensitivity curves of BBO and DECIGO. This result indicates that a possible detection of continuous GWs originated from these objects would corroborate the WD pulsar model.
\end{abstract}



\section{Introduction}
\label{intro}

Over the last decade, there has been an increasing interest of the astrophysics community on highly magnetized white dwarfs (HMWDs) both from the theoretical and observational points of view. These sources constitute at least 10\% of the white dwarfs (WDs) if observational biases are considered~\citep{2007ApJ...654..499K}. These WDs with surface magnetic fields raging from $10^6$~G to $10^9$~G have  been confirmed by the recent results of the Sloan Digital Sky Survey (SDSS) \citep{2009A&A...506.1341K,2010AIPC.1273...19K,2013MNRAS.429.2934K,2015MNRAS.446.4078K}. Besides their high magnetic fields, most of them have been shown to be massive, and responsible for the high-mass peak at $1~\textrm{M}_\odot$ of the WD mass distribution; for instance: REJ 0317--853 has $M \approx 1.35~\textrm{M}_\odot$ and $B\approx (1.7$--$6.6)\times 10^8$~G \citep{1995MNRAS.277..971B,2010A&A...524A..36K}; PG 1658+441
has $M \approx 1.31~\textrm{M}_\odot$ and $B\approx 2.3\times 10^6$~G \citep{1983ApJ...264..262L,1992ApJ...394..603S};
and PG 1031+234 has the highest magnetic field $B\approx 10^9$~G \citep{1986ApJ...309..218S,2009A&A...506.1341K}. The existence of ultra-massive WDs has been revealed in several studies~\citep{2005A&A...441..689A,2007A&A...465..249A,2013MNRAS.430...50C,2013ApJ...771L...2H,2017MNRAS.468..239C,2019A&A...625A..87C,10.1093/mnras/sty3016,2018MNRAS.480.4505J}.

Typically, WDs rotate with periods of days or even years. Recently, a WD pulsar known as AR Scorpii was discovered with a period of $1.97$ min, emitting radiation in a broad range of frequencies, typically of neutron star (NS) pulsars \citep{2016Natur.537..374M}. The spindown power is an order of magnitude larger than the observed luminosity (dominated by the X-rays), which, together with an absence of obvious signs of accretion, suggests that AR Sco is primarily rotation-powered. The AR Sco’s broadband spectrum is characteristic of synchrotron radiation, requiring relativistic electrons, possibly originated from the neighborhood of the WD and accelerated to almost the speed of light~\citep{2017JPhCS.861a2005L}. Furthermore, other sources have been proposed as candidates of WD pulsars. A specific example is AE Aquarii, the first WD pulsar identified, with a short rotation period of $P=33.08$~s~\citep{2008PASJ...60..387T} and spinning down at a rate $P=5.64\times 10^{-14}$~s/s. The rapid braking of the WD and the nature of hard X-ray pulses detected with SUZAKU space telescope~\citep{2008AdSpR..41..512T} can be explained in terms of spin-powered pulsar mechanism~\citep[see][]{1998A&A...338..521I}. On the other hand, the X-ray Multimirror Mission (XMM) - Newton satellite has observed a WD faster than AE Aquarii. Mereghetti et al.~\citep{2009Sci...325.1222M} showed that the X-ray pulsator RX J0648.0-4418 is a massive WD with mass $M=1.28M_\odot$ and radius $R = 3000$ km~\citep[see][for derived mass-radius relations for massive oxygen-neon WDs that predict this radius]{refId0,Althaus2007}, with a very fast spin period of $P = 13.2$~s, that belongs to the binary system HD 49798/RX J0648.0-4418. More recently, \cite{2020ApJ...898L..40L} report on XMM-Newton observations that reveal CTCV J2056-3014 to be an X-ray-faint intermediate polar harboring an extremely fast-spinning WD with a coherent 29.6 s pulsation.

Notwithstanding, several currently studies of fast-rotating and magnetized WDs have been done, in particular the one involving WD pulsars in an alternative description for Soft Gamma Repeaters (SGRs) and Anomalous X-Ray Pulsars (AXPs) \citep[see][and references therein]{Malheiro/2012,Coelho/2014,2016IJMPD..2541025L,2016JCAP...05..007M, 2017MNRAS.465.4434C}. From this perspective, a canonical spin-powered pulsar model can explain the process of energy emission released by dipole radiation in a WD, since they share quite similar aspects \citep{Usov/1988,Coelho/2014}. In addition, these sources could also be candidates for GW emission, since the huge magnetic field can deform the star in a non-symmetrical way, thus generating a variation in the quadrupolar moment of the star.

In a second proposed scenario, that of a WD pulsar, the optical/IR data are explained by the WD photosphere and by a disk~\citep{2013ApJ...772L..24R}.  Recently, a new  scenario has been  proposed to  explain the spectral energy distribution (SED) of 4U 0142+61, from mid-infrared up to hard X-rays~\citep{Sarah2019}. In this model, the persistent emission comes from an accreting isolated magnetic WD surrounded by a debris disk, having gas and dusty regions. 

On the other hand, direct observations of GWs have recently been made by LIGO and Virgo. The first event was detected in 2015 by LIGO \citep{ABBOTT/2016}. This event, named GW150914, came from the merging of two black holes of masses $ \sim  35.6\,\rm{M}_{\odot}$ and $30.6\,\rm{M}_{\odot}$ that resulted in a black hole of mass $ \sim  63.1\,\rm{M}_{\odot}$. Thereafter, LIGO in collaboration with Virgo observed 9 more such events \citep{ABBOTT/2017a,ABBOTT/2017c,ABBOTT/2017b,Abbott_2019}. In addition, the event GW170817 reports the first detection of GW from a binary NS inspiral \citep{abbott2017d}.
All GW detections are within a frequency band ranging from $10$ Hz to $1000$ Hz, which is the operating band of LIGO and Virgo. As is well known, there are proposed missions for lower frequencies, such as LISA \citep{AMARO/2017,cornish2018}, whose frequency band is of $(10^{-4}-0.01)$ Hz, BBO \citep{harry/2006,yagi2011} and DECIGO \citep{kawamura/2006,yagi2017} in the frequency band ranging from $0.01$ Hz to $10$ Hz.

Different  possibilities  of  generation  of continuous GWs have already been proposed~\citep[see e.g.,][and references therein]{1996A&A...312..675B,2016EPJC...76..481D,2016ApJ...831...35D,2016JCAP...07..023D,10.1093/mnras/stx2119,2017EPJC...77..350D,2017ApJ...844..112G,Schramm/2017,2018EPJC...78..361P,2019arXiv190600774D}. More recently, Kalita and Mukhopadhyay \citep{2019MNRAS.490.2692K} shows that continuous GWs can be emitted from rotating magnetized WDs and will possibly be detected by the upcoming GW detectors such as LISA, DECIGO and BBO. The main goal of the present paper is to extend our previous study \citep{10.1093/mnras/staa205} in which we investigated the gravitational radiation from three fast-spinning magnetized WDs, which have high rotations (a few seconds to minutes) and  huge magnetic fields ($10^{6}$ G to $\sim 10^{9}$ G), considering two emission mechanisms: matter accretion and magnetic deformation. In both cases, the GW emission is generated by asymmetry around the rotation axis of the star due to accumulated mass in the magnetic poles and due to the intense magnetic field, respectively. Here, we explore the magnetic deformation mechanism of gravitational radiation emission in SGRs/AXPs as fast-spinning magnetized WD.  

This paper is organized as follows. In Sec.~\ref{sec:5}, we present the aspects
of the model used to explain SGRs/AXPs. In Sec.~\ref{sec:6}, we describe the mechanism of GW emission by deriving the equations for the gravitational amplitude and luminosity. In Sec.~\ref{sec:9} we present and discuss the calculations applied to SGRs/AXPs described as WD pulsars. Finally, in Sec.~\ref{sec:13} we summarize the main conclusions and remarks.

\section{SGRs/AXPs as white dwarf pulsars }
\label{sec:5}

SGRs and AXPs are a special class of pulsars that present distinct characteristics from radio pulsars and X-ray pulsars (see Table \ref{tabla:SGR/AXP}). They are described by the magnetar model, where they are considered strongly magnetized NS with magnetic field of the order of $10^{12} - 10^{15}$ G. Also, these objects are known as very slow rotating pulsars comparing to ordinary pulsars, with rotational periods in the range of $P \sim 2-12$ s and a high spindown rate of $\dot{P} \sim 10^{-13} - 10^{-10}$~s/s~\citep[see][and references therein]{Olausen/2014}\footnote{For information  about the SGRs/AXPs, we refer the reader to the McGill University's online catalog available at: \url{http://www.physics.mcgill.ca/~pulsar/magnetar/main.html}}.

\begin{table*}
\begin{center}
\caption{Observational quantities taken from McGill Pulsar Group's online catalog for confirmed SGRs/AXPs: Period ($P$). spindown ($\dot{P}$), observed luminosity ($L_{X}$) and distance to the source ($r$).}
\vskip0.3cm
\begin{tabular}{lcccc}
\hline
\textbf{SGR/AXP} & \begin{tabular}[c]{@{}c@{}}$P$\\ (s)\end{tabular} & \begin{tabular}[c]{@{}c@{}}$\dot{P}$\\ ($10^{-11}$ s/s)\end{tabular} & \begin{tabular}[c]{@{}c@{}}$L_{X}$\\ ($10^{33}$ erg/s)\end{tabular} & \begin{tabular}[c]{@{}c@{}}$r$\\ (kpc)\end{tabular} \\ \hline
CXOU J010043.1-721134 & 8.020392 & 1.88 & 65 & 62.4 \\
4U 0142+61 & 8.688692 & 0.2022 & 105 & 3.6 \\
SGR 0418+5729 & 9.078388 & 0.0004 & 0.00096 & 2 \\
SGR 0501+4516 & 5.76207 & 0.594 & 0.81 & 2 \\
SGR 0526-66 & 8.0544 & 3.8 & 189 & 53.6 \\
1E 1048.1-5937 & 6.457875 & 2.25 & 49 & 9 \\
1E 1547.0-5408 & 2.072126 & 4.77 & 1.3 & 4.5 \\
PSR J1622-4950 & 4.3261 & 1.7 & 0.44 & 9 \\
SGR 1627-41 & 2.594578 & 1.9 & 3.6 & 11 \\
CXOU J164710.2-455216 & 10.61064 & $\leqslant$ 0.04 & 0.45 & 3.9 \\
1RXS J170849.0-400910 & 11.00502 & 1.9455 & 42 & 3.8 \\
CXOU J171405.7-381031 & 3.825352 & 6.40 & 56 & 13.2 \\
SGR J1745-2900 & 3.763638 & 1.385 & $\leqslant$ 0.11 & 8.3 \\
SGR 1806-20 & 7.54773 & 49.5 & 163 & 8.7 \\
XTE J1810-197 & 5.540354 & 0.777 & 0.043 & 3.5 \\
Swift J1822.3-1606 & 8.437721 & 0.0021 & $\leqslant$ 0.00040 & 1.6 \\
SGR 1833-0832 & 7.565408 & 0.35 & ... & $\leqslant$ 10 $^{a}$ \\
Swift J1834.9-0846 & 2.482302 & 0.796 & $\leqslant$ 0.0084 & 4.2 \\
1E 1841-045 & 11.78898 & 4.092 & 184 & 8.5 \\
J185246.6+003317 & 11.55871 & $\leqslant$ 0.014 & $\leqslant$ 0.0060 & 7.1 \\
SGR 1900+14 & 5.19987 & 9.2 & 90 & 12.5 \\
SGR 1935+2154 & 3.245065 & 1.43 & ... & $\leqslant$ 10 $^{b}$ \\
1E 2259+586 & 6.979043 & 0.0483 & 17 & 3.2 \\ \hline
\end{tabular}

$^{a}$see \cite{esposito2011}. $^{b}$ see \cite{kozlova2016}.

\label{tabla:SGR/AXP}
\end{center}

\end{table*}

Recently, three SGRs with low magnetic field ($B\sim 10^{12}-10^{13}$~G) have been observed, namely, SGR 0418+5729, Swift J1822.3-1606 and 3XMM J185246.6+00331. These new discoveries open the question concerning the nature of SGRs/AXPs, emerging alternative scenarios, in particular the WD pulsar model. These astronomical observations have based an alternative description of the SGRs/AXPs, which are modeled as rotating highly magnetized and very massive WDs~\citep[see][for further details]{Malheiro/2012,Coelho/2014,2017MNRAS.465.4434C}. From this perspective, a canonical spin-powered pulsar model can explain the process of energy emission released by a dipole radiation in a WD, since they share quite similar aspects~\citep[see][]{Usov/1988}. In this new description, several observational properties are explained as a consequence of the large radius of a massive WD that manifests a new scale of mass density, moment of inertia, rotational energy, and magnetic dipole moment in comparison with the case of NSs~\citep[see e.g.,][and references therein]{Coelho/2014,2016IJMPD..2541025L}.

In the canonical pulsar model, a rotating star with magnetic dipole moment misaligned to the axis of rotation converts rotational energy into electromagnetic energy. Thus, the system emits radiation due to the variation of the magnetic dipole and the pulsar rotation becomes slower. If we consider that all rotational energy loss is converted to electromagnetic energy, we can infer the magnetic field strength on the star's surface, $B_{s}$, as a function of the period $P = 1/f_{rot}$ and its derivative $\dot{P} = dP/dt$ \citep[see, e.g.,][]{Coelho/2014}

\begin{equation}
    B_{s}\sin \phi =  \left ( \frac{3c^{3} I}{8\pi^{2} R^{6} }P \dot{P}{ } \right )^{1/2},
    \label{CMs}
 \end{equation}

\noindent where $I$ is the moment of inertia, $R$ is the star radius, $\phi$ is the unknown angle between the rotation and magnetic dipole axes and $c$ is the speed of light.

Note that because the moment of inertia values for a NS and a WD are different, the magnetic fields required in each model are also different. For example, for a NS with mass $M = 1.4 M_{\odot }$ and radius $R = 10^{6}$ cm, the magnetic field on the star's surface are in the range of $10^{13} - 10^{15}$ G. For a very massive WD with mass $M = 1.4 M_{\odot }$ and radius $R \sim 10^{8}$~cm \citep[see, e.g,][]{Coelho/2014}, the magnetic field has smaller values and is in a range around $10^{9} - 10^{11}$ G, comparable to the inferred values of known HMWDs \citep{2009A&A...506.1341K,2010AIPC.1273...19K,2013MNRAS.429.2934K,2015MNRAS.446.4078K}. 
Thus, these values of mass and radius generate a moment of inertia $I\simeq 1.1\times 10^{49}$ $\rm g.cm^2$. These results clearly show that the scale of the magnetic field in WD is $10^4$ times smaller than for NSs.

In addition, since the high magnetic field can deform the star in a non-symmetrical way, new values for the  ellipticity are expected and, consequently, new values for the GW amplitude as well (see section \ref{sec:6}). 
In the next section, we describe the magnetic deformation mechanism and deduce the GW amplitude and luminosity emitted in this process.

\section{Magnetic deformation mechanism: basic equations}
\label{sec:6}

WDs might generate GWs whether they are not perfectly symmetric around their rotation axes. This asymmetry can occur, for example, due to the huge dipole magnetic field that can make the star become oblate \citep[see, e,g,][]{chandrasekhar1953s}. In this work, we analyze the emission of gravitational radiation from SGRs/AXPs as fast-spinning magnetized WDs by this mechanism.

Thus, in this section we consider the deformation of the WDs induced by their own huge magnetic fields. Due to the combination of magnetic field and rotation, a WD can become  triaxial, presenting therefore a triaxial moment of inertia. In order to investigate the effect arising from the magnetic stress on the equilibrium configuration of the stars, let us  introduce  the equatorial ellipticity, defined as follows \citep[see, e.g.,][]{shapiro/2008,maggiore/2008}

\begin{equation}
  \epsilon = \frac{I_{1} - I_{2}}{I_{3}}.
 \label{elipticidade}
\end{equation}

\noindent where $I_{1}$, $I_{2}$ and $I_{3}$ are main moments of inertia with respect to the ($x$, $y$, $z$) axes, respectively.

If the star rotates around the $z-$axis, then it will emit monochromatic GWs with a frequency twice the rotation frequency, $f_{rot}$, and amplitude given by \citep[see, e.g.,][]{shapiro/2008,maggiore/2008}

\begin{equation}
   h_{df} = \frac{16 \pi^{2} G}{c^{4}}  \frac{I_{3} f_{rot}^{2}}{r} \, \epsilon,
\label{Ampdef2}
\end{equation}

\noindent and the rotational energy of the star decreases at a rate given by \cite[see, e.g.,][]{shapiro/2008,maggiore/2008}

 \begin{equation}
   L_{GW_{df}} = - \frac{2^{11}  \pi^{6} }{5} \frac{G}{c^{5}} I_{3}^{2} \epsilon^{2} f_{rot}^{6}.
\label{Lumidef1}
\end{equation}

On the other hand, recall that the ellipticity of magnetic origin can be written as follows~\citep[see, e.g.,][]{1996A&A...312..675B,2000A&A...356..234K,2006A&A...447....1R}

\begin{equation}
\epsilon = \kappa \frac{B_{s}^{2} R^{4}}{G M^{2}} \sin^2{\phi},
    \label{excentridade}
\end{equation}

\noindent where, as before, $B_{s}$ is the magnetic field strength on the star’s surface, $R$ and $M$ are, respectively, the radius and mass of the star, $\phi$ is the angle between the rotation and magnetic axes, whereas $\kappa$ is the distortion parameter, which depends on the magnetic field configuration and equation of state (EoS) of the star. It is worth mentioning that as the magnetic field of the SGRs/AXPs is inferred from the spindown rate of the star, where it is considered that all rotational energy loss is converted into electromagnetic energy [see Eq.~(\ref{CMs})], care must be taken when using these field values to calculate an additional GW spin-down torque. In fact, in the next section, we show that this procedure is safe and that only a small fraction of the total energy loss goes to GWs.

To proceed, substituting this last equation into Eqs.~(\ref{Ampdef2}) and (\ref{Lumidef1}) and considering $I_3 = 2MR^{2}/5 $, one immediately obtains that

\begin{equation}
   h_{df} = \frac{32 \pi^{2} }{5 c^{4}}  \frac{R^{6} f_{rot}^{2}}{r M} \kappa (B_{s}\sin{\phi})^{2},
\label{Ampdef3}
\end{equation}
and 
\begin{equation}
   L_{GW_{df}} = - \frac{2^{13} \pi^{6} }{5^3 c^{5}} \frac{ R^{12} f_{rot}^{6}}{G M^{2}}\kappa^2 (B_{s}\sin{\phi})^{4}.
\label{Lumidef2}
\end{equation}

Thereby, we find equations for the gravitational luminosity and the GW amplitude which depend on the rotation frequency and the magnetic field strength.

A very interesting equation can be obtained by substituting Eq.~(\ref{CMs}) into Eq.~(\ref{excentridade}), namely
\begin{equation}
\epsilon = \frac{3}{20 \pi^2}\frac{c^{3}}{G M} P{\dot P}\kappa,
    \label{excentridade2}
\end{equation}

\noindent which is independent of the angle $\phi$.

Consequently, by substituting Eq.~ (\ref{excentridade2}) into 
Eqs.~(\ref{Ampdef3}) and (\ref{Lumidef2}) one obtains

\begin{equation}
   h_{df} = \frac{24}{25}  \frac{R^{2}}{cr} \frac{\dot P}{P}\kappa,
\label{Ampdef4}
\end{equation}
and 
\begin{equation}
   L_{GW_{df}} = - \frac{2^9 3^2 \pi^2 }{5^5} \frac{c}{G} R^{4}\left(\frac{\dot P}{P^2}\right)^2\kappa^2.
\label{Lumidef3}
\end{equation}

Therefore, it is not necessary to be concerned about $\phi$ in the calculations of the luminosity and amplitude of GWs. Different values of $\phi$ is only import in the calculation of $B_{s}$. Noticing, however, that $\phi$ and $B_{s}$ are not independent, as can be seen from Eq.~(\ref{CMs}).

Another point to note here is that the GW amplitude (Eq. (\ref{Ampdef4})) depends on the square of the radius. However, it is worth recalling that for compact stars, as WDs, the mass is related to the radius so that the more massive the star is, the smaller the radius will be. Thus, the amplitude implicitly depends on the mass $M$.

One could argue that rotation could be also important in the deformation of the star. In fact, rotation modifies the equatorial radius of the star, as a result the amplitude of the GWs is affected, since $h_{df} \propto R^2$. Thus, the higher the rotation velocity is, the greater is the radius.

Now, we are ready to calculate the GW amplitude and luminosity for SGRs/AXPs as massive fast-spinning WDs. The next section is devoted to this issue as well as the corresponding discussion of the results.

\section{Results and Discussions}
\label{sec:9}

Here we consider that SGRs/AXPs are fast-spinning and magnetized WDs which emit GWs due to the deformation caused by their own intense magnetic field. For this study, we use the magnetic field values inferred from the canonical pulsar model (see Eq. (\ref{CMs})), where it is considered that all the spindown luminosity of the star is converted to electromagnetic luminosity.

Thus, using Eq. (\ref{Lumidef2}), we calculate the GW luminosity for several SGRs/AXPs, considering these objects as a very massive WD of $M_{WD} = 1.4 M_{\odot }$ and radius $R_{WD} = 1.0 \times 10^{8}$ cm \citep{2013ApJ...762..117B}. We adopted $\kappa\simeq 10$, which is a conservative value that holds for an incompressible fluid star with a dipole magnetic field \citep[see, e.g.][]{1954ApJ...119..407F}. Similar values are obtained when relativistic models based on a polytropic EoS with dipole magnetic field are considered  \citep[see, e.g.][]{2000A&A...356..234K}. In fact, this parameter can take different values depending on the EoS and the geometry of the magnetic field \citep[see, e.g.][for a usefull discussion regarding this issue]{2006A&A...447....1R}.

The result of this calculation is presented in Table \ref{tabla:efici_def_SGR}, which also displays the ellipticity $\epsilon$, the spindown luminosity $L_{sd}$ ($ = 4\pi^2 I_3 f_{rot} \dot{f}_{rot} $) and the efficiency $\eta_ {df}$ ($=L_{GW_{df}}/L_{sd}$). Notice that the efficiencies are around $10^{-9}$ to $10^{-13}$. This implies that the gravitational luminosity is much smaller than the spindown luminosity when considering the magnetic fields inferred by the dipole model. Thus, we see that we can apply these magnetic field values to calculate the  GW amplitude, since the emission of gravitational energy is negligible as compared to the rotational energy loss rate, not changing significantly the inferred magnetic fields.

\begin{table*}
\begin{center}
\caption{ Parameters for SGRs/AXPs based on a very massive white dwarf of fiducial parameters, $M_{WD} = 1.4 M_{\odot }$ and radius $R_{WD} = 1.0 \times 10^{8}$ cm~\citep{2013ApJ...762..117B}. We adopt in the calculation $\kappa = 10$.}
\begin{tabular}{lccclc}
\hline
\multicolumn{6}{c}{\textbf{SGRs/AXPs  as massive WD of $M_{WD} = 1.4 M_{\odot }$}} \\ \hline
\textbf{SGRs/AXPs} & \begin{tabular}[c]{@{}c@{}}$B_{s}\sin{\phi}$ \\ ($10^{10}$ G)\end{tabular} & \begin{tabular}[c]{@{}c@{}}$\epsilon$\\ ($10^{-6}$)\end{tabular} & \begin{tabular}[c]{@{}c@{}}$L_{sd}$ \\ ($10^{37}$ erg/s)\end{tabular} & \multicolumn{1}{c}{\begin{tabular}[c]{@{}c@{}}$L_{GW_{df}}$ \\ (erg/s)\end{tabular}} & \begin{tabular}[c]{@{}c@{}}$\eta _{df}$\\ ($10^{-10}$)\end{tabular} \\ \hline
CXOU J010043.1-721134 & 4.12 & 3.28 & 1.59 & \multicolumn{1}{c}{5.48 $\times 10^{27}$} & 3.43 \\
4U 0142+61 & 1.41 & 0.382 & 0.135 & \multicolumn{1}{c}{4.59 $\times 10^{25}$} & 0.340 \\
SGR 0418+5729 & 0.0640 & 0.000790 & 0.000234 & \multicolumn{1}{c}{1.51 $\times 10^{20}$} & 0.000645 \\
SGR 0501+4516 & 1.96 & 0.745 & 1.36 & 2.05 $\times 10^{27}$ & 1.51 \\
SGR 0526-66 & 5.87 & 6.66 & 3.19 & 2.20 $\times 10^{28}$ & 6.91 \\
1E 1048.1-5937 & 4.05 & 3.16 & 3.66 & 1.87 $\times 10^{28}$ & 5.10 \\
1E 1547.0-5408 & 3.34 & 2.15 & 235.0 & 7.92 $\times 10^{30}$ & 33.7 \\
PSR J1622-4950 & 2.89 & 1.60 & 9.20 & 5.30 $\times 10^{28}$ & 5.75 \\
SGR 1627-41 & 2.36 & 1.07 & 47.7 & 5.11 $\times 10^{29}$ & 10.7 \\
CXOU J164710.2-455216 & 0.692 & 0.0924 & 0.0147 & 8.10 $\times 10^{23}$ & 0.0552 \\
1RXS J170849.0-400910 & 4.92 & 4.67 & 0.641 & 1.66 $\times 10^{27}$ & 2.59 \\
CXOU J171405.7-381031 & 5.25 & 5.33 & 50.1 & 1.23 $\times 10^{30}$ & 24.5 \\
SGR J1745-2900 & 2.43 & 1.14 & 11.4 & 6.18 $\times 10^{28}$ & 5.41 \\
SGR 1806-20 & 20.5 & 81.3 & 50.4 & 4.84 $\times 10^{30}$ & 96.0 \\
XTE J1810-197 & 2.20 & 0.937 & 2.00 & 4.11 $\times 10^{27}$ & 2.05 \\
Swift J1822.3-1606 & 0.141 & 0.00386 & 0.00153 & 5.58 $\times 10^{21}$ & 0.00365 \\
SGR 1833-0832 & 1.73 & 0.576 & 0.354 & 2.40 $\times 10^{26}$ & 0.677 \\
Swift J1834.9-0846 & 1.49 & 0.430 & 22.8 & 1.07 $\times 10^{29}$ & 4.70 \\
1E 1841-045 & 7.37 & 10.5 & 1.09 & 5.56 $\times 10^{27}$ & 5.08 \\
3XMM J185246.6+003317 & 0.427 & 0.0352 & 0.00397 & 7.05 $\times 10^{22}$ & 0.0177 \\
SGR 1900+14 & 7.34 & 10.4 & 28.7 & 7.43 $\times 10^{29}$ & 25.9 \\
SGR 1935+2154 & 2.29 & 1.01 & 18.3 & 1.18 $\times 10^{29}$ & 6.45 \\
1E 2259+586 & 0.617 & 0.0735 & 0.0624 & 6.34 $\times 10^{24}$ & 0.102 \\ \hline
\end{tabular}
\label{tabla:efici_def_SGR}
\vskip0.3cm

\end{center}
\end{table*}

It is worth recalling that the rotation periods as short as the ones observed in SGRs/AXPs, which ranges from, $P\sim 2-12$~s, can be really attained by WDs. \citet{2013ApJ...762..117B} showed that the range of minimum rotation periods of massive WDs is of $0.3\leq P\leq 2.2$~s, depending on the nuclear composition~\footnote{The relatively long minimum period of $^{56}$Fe spinning WDs,$\sim2.2$~s, implies that spinning WDs describing SGRs/AXPs have to be composed of nuclear compositions lighter than $^{56}$Fe, e.g., $^{12}$C or $^{16}$O~\citep[see][for details]{2013ApJ...762..117B}}. We refer the reader to the paper of these very authors for further discussion and details.

\begin{figure}
\centering
\includegraphics[width=8.9cm, height= 6.5cm]{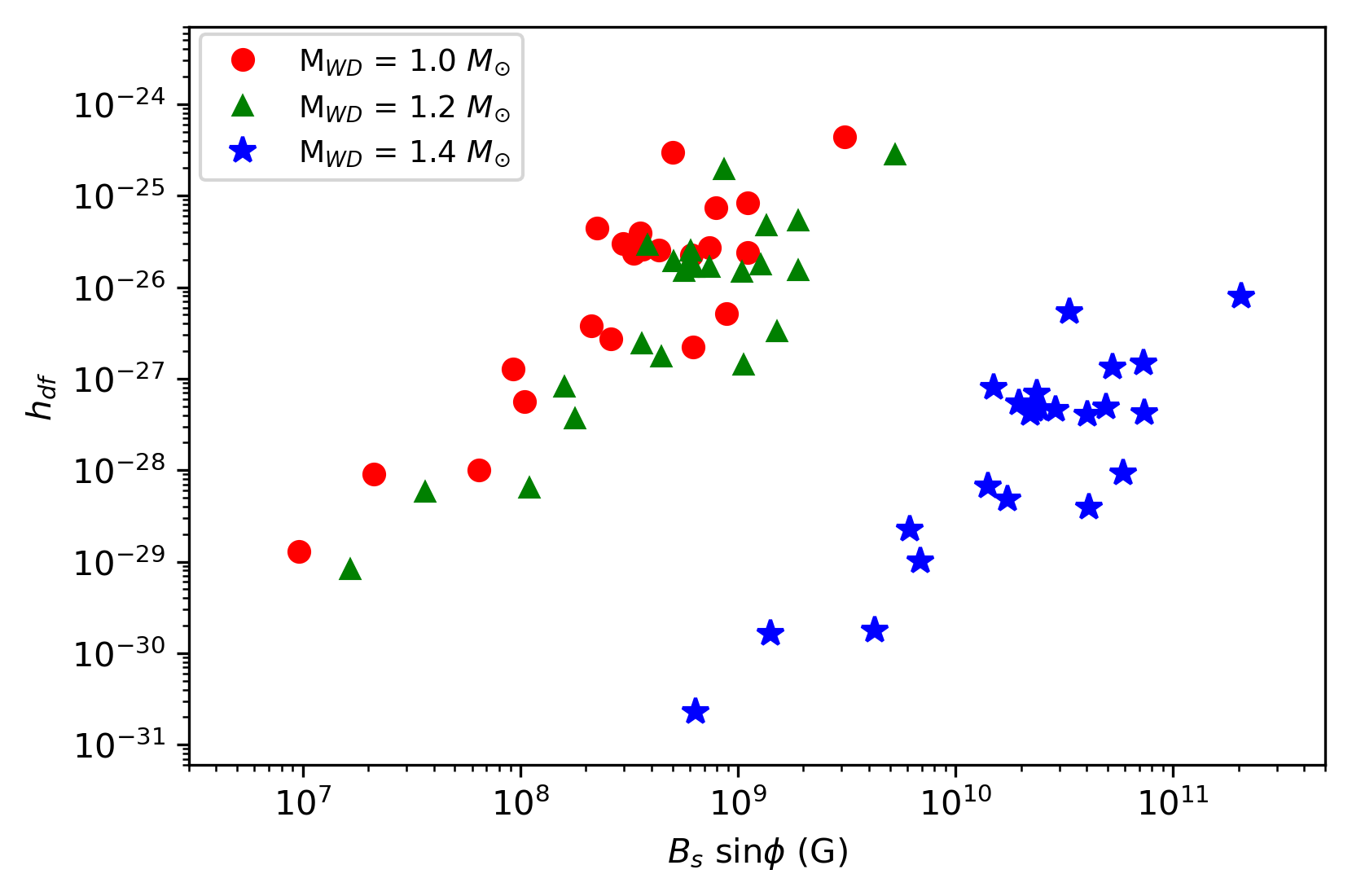} 
\caption{GW amplitude as function of the magnetic field for SGRs/AXPs as fast-spinning and magnetized WDs.}
\label{fig:1}       
\end{figure}

Figure \ref{fig:1} shows the GW amplitudes versus the magnetic fields for the 23 confirmed SGRs/AXPs (see also Table \ref{tabla:SGR/AXP}). Besides our fiducial WD model ($M_{WD} = 1.4 M_{\odot }$ of $R_{WD} = 1.0 \times 10^{8}$ cm), we consider two additional models, namely,
1.2 $M_{\odot }$ [$R_{WD} = 6.0 \times 10^{8}$ cm] and 1.0 $M_{\odot }$ [$R_{WD} = 7.5 \times 10^{8}$ cm] \citep[see][for further details about the mass-radius relation]{Boshkayev2013}. We stress that the stability of rotating WDs was analyzed taking into account the mass-shedding limit, inverse $\beta$-decay, and pycnonuclear instabilities \citep[see also][for several macro and micro instabilities in WDs]{2014ApJ...794...86C}, as well as the secular axisymmetric instability \citep[see Fig.1 of][for details]{Boshkayev2013}.
Note that for a given source the predicted magnetic field (GW amplitude) can vary almost two orders of magnitude depending on the assumed parameters.

\begin{figure}
\centering
\includegraphics[width=9.2cm, height= 6.9cm]{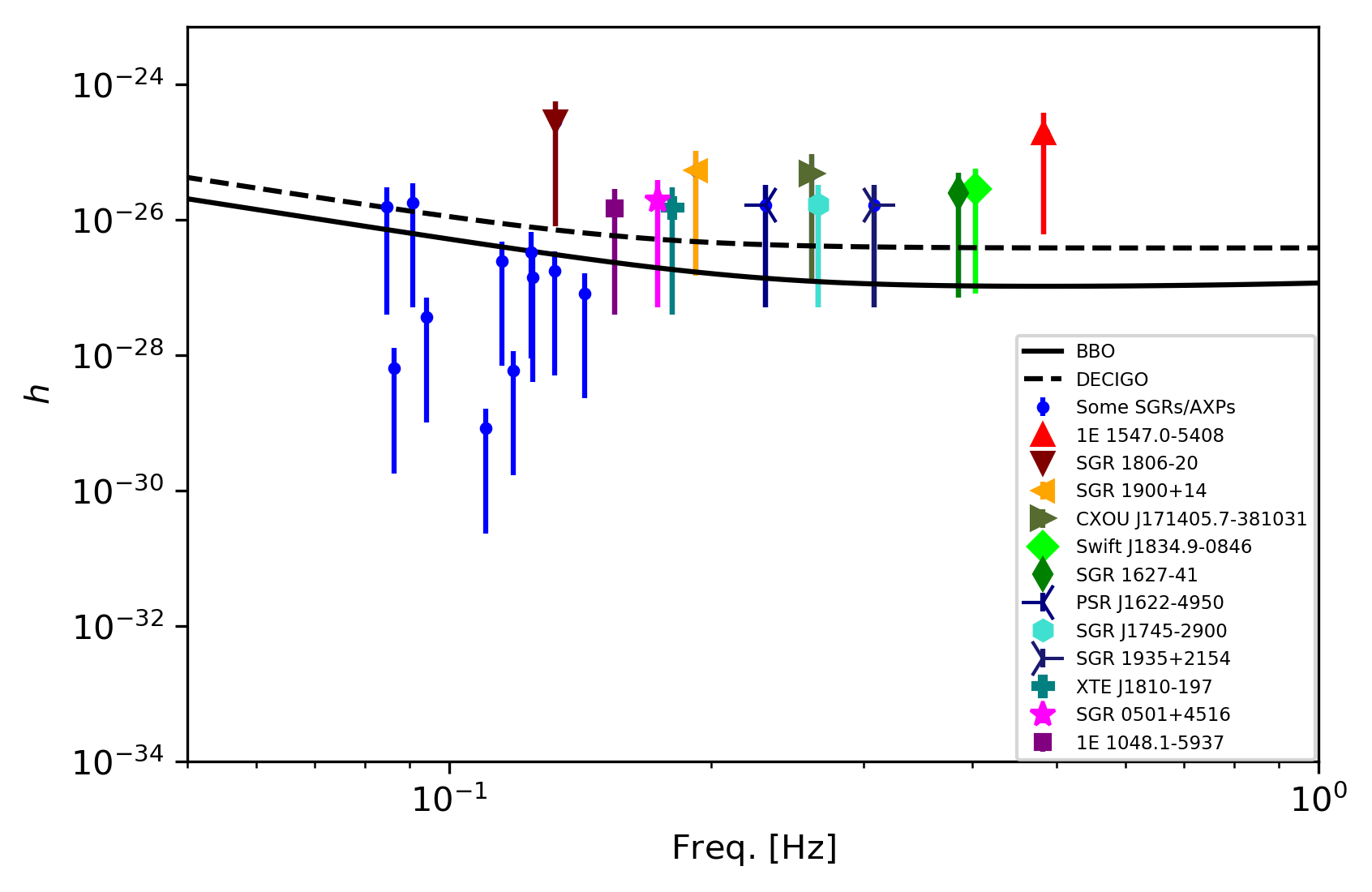} 
\caption{GW amplitude as a function of frequency for SGRs/AXPs as fast-spinning and magnetized WDs for masses in the interval  $1.0 M_{\odot} \leq  M_{WD} \leq  1.4 M_{\odot}$ (or $7.5 \times 10^{8}$ cm $\geq  R_{WD} \geq 1.0 \times 10^{8}$ cm), represented by the vertical bars, from top to bottom. The bullets stand for $M_{WD} = 1.2 M_{\odot }$ ($R_{WD} = 6.0 \times 10^{8}$ cm). Also plotted the sensitivity curves for BBO and DECIGO for $SNR = 8$ and integration time $T = 1$ year.}
\label{fig:2}       
\end{figure}

Figure \ref{fig:2} shows the GW amplitude as a function of frequency for some SGRs/AXPs, where the bullets stands for $M_{WD} = 1.2 M_{\odot }$ and the vertical bars, that crosses the bullets, stands for  $1.0 M_{\odot} \leq   M_{WD} \leq  1.4 M_{\odot} $, from top to bottom.
We also plot the sensitivity curves for BBO and DECIGO. It is worth mentioning that to plot the sensitivity curves, we use the minimum amplitude, $h_{min}$, that can be measured by the detector, for a periodic signal, for a given signal-to-noise ratio (SNR) and observation time $T$ \citep[see, e.g.,][for further details]{maggiore/2008}. Thereby, Fig. \ref{fig:2} presents the GW amplitudes for the sources ($h_{df}$) and the sensitivity curves are set to SNR = 8 and $T = 1$ year. Note that we do not display the sensitivity curve for LISA because these sources are far below it. 

Notice that some SGRs/AXPs produce GWs with amplitudes that can be detected by BBO and DECIGO. For example, 1E 1547.0-5408 and SGR 1806-20 could well be detected for the entire mass range considered. SGR 1900+14 and CXOU J171405.7-381031, in turn, are detectable for the entire mass range only by BBO. For these sources to be observed by DECIGO, they must have mass $M_{WD} \lesssim 1.3 M_{\odot }$. The sources Swift J1834.9-0846, SGR 1627-41, PSR J1622-4950, SGR J1745-2900 and SGR 1935+2154 could be observed by BBO and DECIGO if they have mass $M_{WD} \lesssim 1.3 M_{\odot }$ and $M_{WD} \lesssim 1.2 M_{\odot }$, respectively. However, if these SGRs/AXPs have $M_{WD} \sim 1.4 M_{\odot }$, they should be detectable only by BBO and with an integration time of $T = 5$ years. XTE J1810-197, SGR 0501+4516 and 1E 1048.1-5937 could also be seen by BBO and DECIGO if these objects have $M_{WD} \lesssim 1.3 M_{\odot }$ and $M_{WD} \lesssim 1.2 M_{\odot }$, respectively, but they will not be observed if they have mass $M_{WD} \sim 1.4 M_{\odot }$, even considering  $T = 5$ years.

\begin{figure}
\centering
\includegraphics[width=9.2cm, height= 6.7cm]{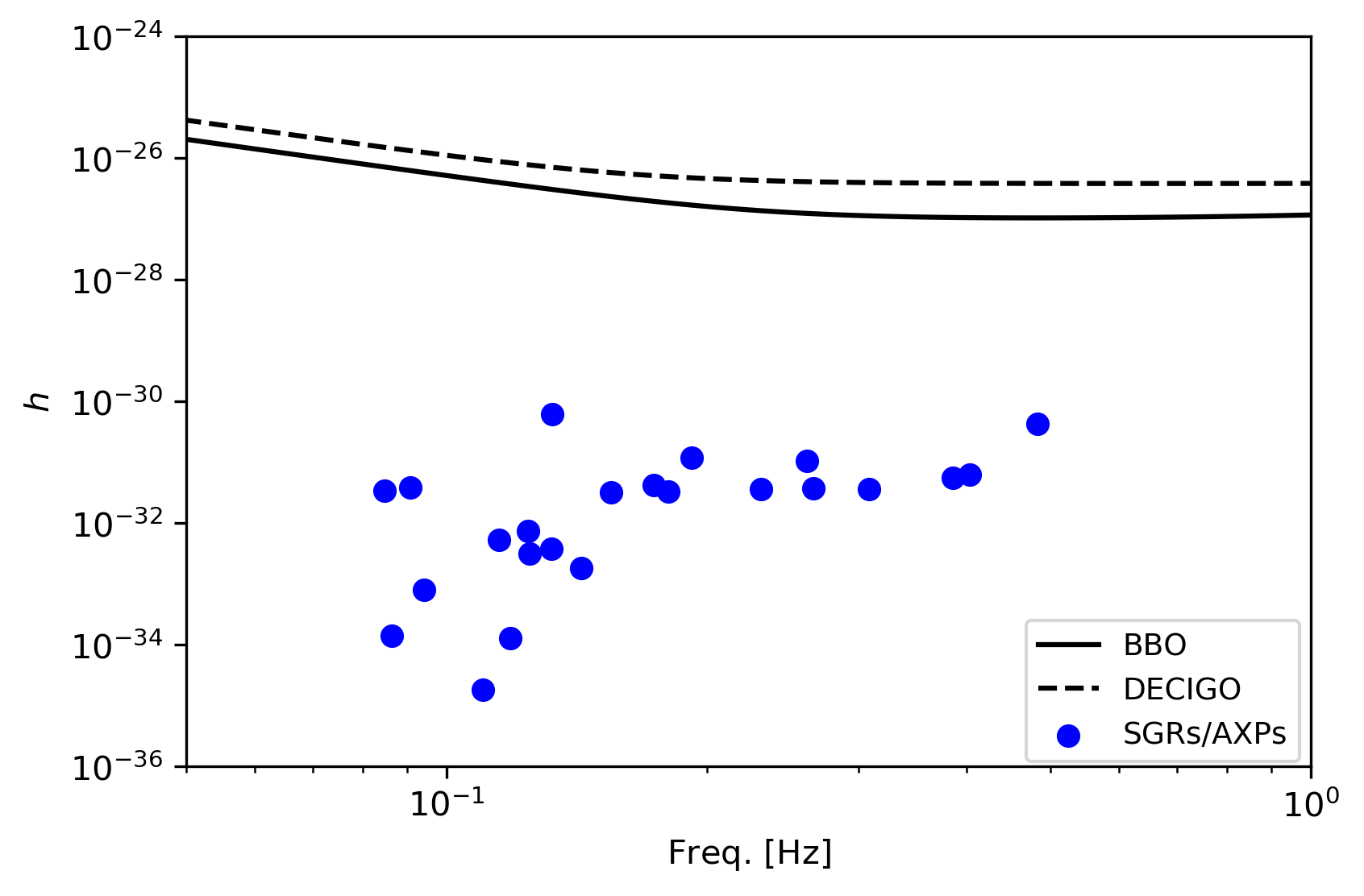} 
\caption{ GW amplitude as a function of frequency for SGRs/AXPs as NSs. Also plotted the sensitivity curves for BBO and DECIGO for $SNR = 8$ and integration time of $T = 1$ year. We consider a NS of $M = 1.4 M_{\odot}$, radius $R = 10$ km and ellipticity given by  $\epsilon = 10 B^2 R^4 \sin^2{\phi} / G M^2$ from \citet{2017EPJC...77..350D}. }
\label{fig:3}       
\end{figure}

Therefore, SGRs/AXPs described as WDs, which have moments of inertia four orders of magnitude greater than a NS, would generate GW amplitudes much larger than SGRs/AXPs described as NSs (see Figure \ref{fig:3}). Consequently, if these sources are NSs the GW amplitudes generated are far below the sensitivity curves of BBO and DECIGO. Thus, if these space based instruments observe continuous GWs from these SGRs/AXPs, this would corroborate the model of fast-spinning and magnetic WDs. This supports the description of SGR and AXPs as belonging to a class of very fast and magnetic massive WDs in perfect accord with recent astronomical observations of HMWDs.

It is worth noticing that from Eq. (\ref{Ampdef4}) that $h_{df} \propto \kappa$. Thus, if we consider an even more conservative value for $\kappa$, e.g., of the order of the unit, the GW amplitude decreases an order of magnitude and some sources in Fig. \ref{fig:2} will be below the sensitivity curves of BBO and DECIGO. XTE J1810-197, SGR 0501 4516 and 1E 1048.1-5937, for example, will be not detected with this $\kappa$ value, whereas the sources 1E 1547.0-5408 and SGR 1806-20 must have mass $M_{WD} \lesssim 1.3 M_{\odot }$ to be seen by the two spacial detectors. As for the other detectable sources in Fig. \ref{fig:2}, they cannot be too massive ($\gtrsim$ $1.2 M_{\odot }$) to continue being observed by BBO.

\section{Summary}
\label{sec:13}

Besides the search and detection of GWs from the merger events \citep{Abbott_2019,abbott2017d}, the search for continuous GWs has been of great interest in the  scientific community. It is well known that, besides compact binaries, rapidly rotating NSs are promising sources of GWs which could be detected in a near future by Advanced LIGO (aLIGO) and Advanced Virgo (AdV), and also by the planned Einstein Telescope (ET) and the space-based LISA, BBO and DECIGO. These sources generate continuous GWs whether they are not perfectly symmetric around their rotation axis, i.e. if they present some equatorial ellipticity. Undoubtedly, SGRs and AXPs are also good candidates in this context. Here we investigate the gravitational radiation from these objects described as fast-spinning WDs using the magnetic deformation mechanism. It is worth stressing that these putative uncommon WDs are known to have a high rotation (a few seconds to minutes) and a huge magnetic field ($10^{6}$ G to $\sim 10^{10}$ G).

Then, by describing SGRs/AXPs as rotation-powered WD pulsars, we consider the role played by the magnetic dipole field on the deformation of these objects and its consequences as regards the generation of GWs considering a mass range $1.0 M_{\odot} \leq   M_{WD} \leq  1.4 M_{\odot}$ for these sources. It is worth mentioning that, this is the first time in the literature that the GW counterpart for SGRs/AXPs are modeled as fast-spinning and magnetized WDs.

We note that some SGRs/AXPs, described as fast-spinning and massive WDs, emit GWs with amplitudes that could be detected by BBO and DECIGO, namely: 1E 1547.0-5408 and SGR 1806-20 can be observed for the entire considered mass range for one year of observation time, while SGR 1900+14, CXOU J171405.7-381031, Swift J1834.9-0846, SGR 1627-41, PSR J1622-4950, SGR J1745-2900 and SGR 1935+2154 can be detected for the entire considered mass range for five years of observation time. The sources XTE J1810-197, SGR 0501+4516 and 1E 1048.1-5937, in turn, can also be observed if they do not have so large masses.

Last but not least, it is worth mentioning that recent astronomical observations suggest that we should revisit the real nature of AXP/SGRs: are they really magnetars or fast-spinning and magnetized WDs? Thereby, a possible detection of continuous GWs coming from SGRs/AXPs would be a good indication that could corroborate the WD model, because for the NSs description, they are far below the BBO and DECIGO sensitivity curves.  We also encourage  future observational campaigns to determine the radii and the magnetic field of these sources to elucidate the real nature of SGRs and AXPs.

\section*{Acknowledgements}
We thank the anonymous referee for useful criticisms and suggestions which have improved our paper.
M.F.S. thanks CAPES for the financial support. J.G.C. is likewise grateful to the support of CNPq (421265/2018-3 and 305369/2018-0). J.C.N.A. thanks FAPESP (2013/26258-4) and CNPq (308367/2019-7) for partial financial support.

\section*{Data Availability}

The data underlying this article will be shared on reasonable request to the corresponding author.





\bibliographystyle{mnras}
\bibliography{references} 


\bsp	
\label{lastpage}
\end{document}